\documentclass[twocolumn,showpacs,prd]{revtex4}
\usepackage{graphicx}
\usepackage{dcolumn}
\usepackage{bm}
\begin{document}

\title{Bremsstrahlung Energy Losses for Cosmic Ray Electrons and Positrons}
\author{A. Widom and J. Swain}
\affiliation{Physics Department, Northeastern University, Boston MA USA}
\author{Y.N. Srivastava}
\affiliation{Physics Department, University of Perugia, Perugia IT}

\begin{abstract}
Recently cosmic ray electrons and positrons, i.e. cosmic ray charged leptons, 
have been observed. To understand the distances from our solar system to the 
sources of such lepton cosmic rays, it is important to understand energy losses 
from cosmic electrodynamic fields. Energy losses for ultra-relativistic electrons 
and/or positrons due to classical electrodynamic bremsstrahlung are computed. 
The energy losses considered are (i) due to Thompson scattering from fluctuating 
electromagnetic fields in the background cosmic thermal black body radiation 
and (ii) due to the synchrotron radiation losses from quasi-static domains of 
cosmic magnetic fields. For distances to sources of galactic length 
proportions, the lepton cosmic ray energy must be lass than about a TeV. 
\end{abstract}


\maketitle

\section{Introduction \label{intro}}

Measurements of the energy distribution of cosmic rays incident on earth bound 
detectors has previously been restricted to baryons\cite{Stanev:2002,Yao:2004}. 
More recently, experimental observations of electrons and positrons as a charged 
lepton components to cosmic rays have also been observed\cite{Ting:2014}. It is 
thereby of great interest to understand the distances from our solar system to 
possible sources of these charged leptons. The understanding relies on possible 
energy losses due to interactions with cosmic electromagnetic fields. In what 
follows we consider energy losses due to bremsstrahlung radiation from charged 
particle accelerations\cite{Landau:1999}.

These measurements give rise to issues of how much radiative damping of these lepton 
energies exist. Such energy losses would be inportant for estimating how distant may 
be the sources of lepton cosmic rays. For cosmic ultra-relativistic leptons of energy 
\begin{math} {\cal E}=mc^2\gamma \end{math} with 
\begin{math} \gamma \gg 1 \end{math} we estimate in Secs. \ref{crd} and 
\ref{cb} that within the distance scales of our galaxy, energy losses from  
classical bremsstrahlung can be neglected if 
\begin{math} 1 \ll \gamma \ll 10^7 \end{math}.

\section{Classical Radiation Damping \label{crd}}

When a non-relativistic \begin{math} |{\bf v}|\ll c  \end{math} classical 
particle with charge \begin{math} e \end{math}, mass 
\begin{math} m \end{math} and acceleration \begin{math} {\bf a} \end{math} 
moves through a classical electromagnetic field, it radiates electromagnetic 
power \begin{math} {\cal P}  \end{math} determined by  
\begin{equation}
{\cal P}=\frac{2e^2}{3c^3}\overline{|{\bf a}|^2}
=\frac{2e^4}{3m^2c^3}\overline{|{\bf E}|^2}, 
\label{crd1}
\end{equation}
wherein the non-relativistic acceleration 
\begin{math} {\bf a}=(e{\bf E}/m)  \end{math} has been employed. 
The incoming electromagnetic energy flux \begin{math} {\cal S} \end{math} 
of a plane wave incident on a charged particle per unit time per unit area is 
given by  
\begin{equation}
{\cal S}=\frac{c}{4\pi }\overline{|{\bf E}|^2}\ . 
\label{crd2}
\end{equation}
Thompson employed Eqs.(\ref{crd1}) and (\ref{crd2}) to derive the total cross 
section for the classical charged particle to scatter an electromagnetic wave; 
i.e. \begin{math} \sigma = {\cal P}/{\cal S} \end{math} yields
\begin{equation}
\sigma =\left(\frac{8\pi r_c^2}{3}\right) \ \ \ \ {\rm wherein}
\ \ \ \ r_c=\left(\frac{e^2}{mc^2}\right).
\label{crd3}
\end{equation}
Numerically, 
\begin{eqnarray}
\sigma_{\rm electron} \approx 6.652446\times 10^{-25}\ {\rm cm^2}\ , 
\nonumber \\ 
\sigma_{\rm proton} \approx 1.973104\times 10^{-31}\ {\rm cm^2}\ .
\label{crd4}
\end{eqnarray}
To understand the manner in which radiation gives rise to energy losses of 
classical relativistic charged particles it is merely necessary to generalize 
Thompson's above argument to the fully relativistic form of classical electromagnetic 
theory.

\subsection{Relativistic Notation \label{rn}}

The proper time of a moving classical particle will be written as 
\begin{equation}
-c^2d\tau^2 =\eta_{\mu \nu}dx^\mu dx^\nu
\label{rn1}
\end{equation}
with the metric signature \begin{math} (+,+,+,-)  \end{math}. The four 
velocity is then 
\begin{eqnarray}
dx^\mu =(d{\bf r},cdt),
\nonumber \\ 
v^\mu = \frac{dx^\mu}{d\tau}\ \ {\rm and}
\ \ {\bf v}=\frac{d{\bf r}}{dt}\ ,
\nonumber \\ 
v^\mu =(\gamma {\bf v},\gamma c)\ \ {\rm wherein} 
\ \ \gamma =\frac{1}{\sqrt{1-|{\bf v}/c|^2}}\ ,
\label{rn2}
\end{eqnarray}
so that 
\begin{equation}
v^\mu v_\mu =-c^2.
\label{rn3}
\end{equation}
The four acceleration is defined as 
\begin{eqnarray}
w^\mu = \frac{dv^\mu }{d\tau }
\ \ {\rm and} \ \ {\bf a}=\frac{d(\gamma {\bf v})}{dt}, 
\nonumber \\ 
w^\mu = (\gamma {\bf a}, \gamma {\bf v\cdot a}/c), 
\nonumber \\ 
w^\mu w_\mu =\gamma^2 \left[|{\bf a}|^2-
\frac{|{\bf v\cdot a}|^2}{c^2} \right],
\label{rn4}
\end{eqnarray}
so that 
\begin{equation}
v^\mu w_\mu =0.
\label{rn5}
\end{equation}
The four momentum 
\begin{equation}
p^\mu =mv^\mu \ \ \ {\rm obeys}
\ \ \ p^\mu p_\mu =-m^2c^2
\label{rn6}
\end{equation}
in virtue of Eq.(\ref{rn3}).

\subsection{Classical Electrodynamic Fields \label{cef}}

In terms of the vector potential 
\begin{math} A^\mu =({\bf A},\Phi) \end{math},  
the electrodynamic fields are described by 
\begin{equation}
{\bf E}=
-\frac{1}{c}\left(\frac{\partial {\bf A}}{\partial t}\right) 
-{\bf grad}\Phi \ \ \ {\rm and} 
\ \ \ {\bf B}=curl{\bf A}.
\label{cef1}
\end{equation}
Equivalently, as a tensor 
\begin{equation}
F_{\mu \nu}=\partial_\mu A_\nu - \partial_\nu A_\mu . 
\label{cef2}
\end{equation}
The Lorentz force plus the classical radiation reaction force on a 
charge may be written as 
\begin{equation}
mw^\mu =\frac{e}{c} F^{\mu \nu} v_\nu  +{\cal F}^\mu ,  
\label{cef3}
\end{equation}
wherein \begin{math} {\cal F}^\mu \end{math} is the radiation 
damping force,
\begin{eqnarray}
{\cal F}^\mu = -\left(\frac{2e^4}{3m^2c^4}\right) \times 
\nonumber \\ 
\left[F_{\beta }^{\ \mu }F^{\beta \alpha }\left(\frac{v_\alpha }{c}\right)
+ F^{\nu \alpha}\left(\frac{v_\alpha }{c}\right)
F_\nu ^{\ \beta}\left(\frac{v_\beta }{c}\right)\left(\frac{v^\mu }{c}\right)\right].  
\label{cef4}
\end{eqnarray}
It is worthy of note that the damping force may be written in terms 
of the Maxwell electromagnetic energy-pressure tensor,  
\begin{equation}
T_{\mu \nu}=\frac{1}{4\pi}\left[F_{\mu }^{\ \lambda}F_{\nu \lambda }
-\frac{1}{4}\eta_{\mu \nu }
\big(F^{\alpha \beta}F_{\alpha \beta}\big)\right]. 
\label{cef5}
\end{equation}
In virtue of Eqs.(\ref{cef4}) and (\ref{cef5}) we have 
\begin{eqnarray}
r_c=\left(\frac{e^2}{mc^2}\right)\ \ \ {\rm and}
\ \ \ \sigma = \left(\frac{8\pi r_c^2}{3}\right), 
\nonumber \\ 
{\cal F}^\mu = -\sigma \left[T^{\mu \nu}\left(\frac{v_\nu}{c}\right)+
\left(\frac{v_\alpha}{c}T^{\alpha \beta}\frac{v_\beta}{c}
\right)\left(\frac{v^\mu }{c}\right)\right].
\label{cef6}
\end{eqnarray}
Eq.(\ref{cef6}) is the completely relativistic version of the Thompson 
cross section \begin{math} \sigma \end{math} outlined in the above 
Sec.\ref{crd}, Eqs.(\ref{crd1}), (\ref{crd2}) and (\ref{crd3}).
The Thomson cross section \begin{math} \sigma  \end{math} fully determines 
the strength of the classical radiation damping force.

\subsection{Particle Orbits \label{po}}

In virtue of Eq.(\ref{cef3}), the orbits of a charged particle in space is 
determined by the force on a charge 
\begin{equation}
\frac{d{\bf p}}{dt}=m{\bf a}=e\left[{\bf E}+\frac{\bf v \times B}{c}\right]+{\bf f}.
\label{po1}
\end{equation}
wherein \begin{math} {\bf f} \end{math} is the radiation damping retarding 
force determined by Eqs.(\ref{cef3}) and (\ref{cef6}). One requires the Maxwell 
electromagnetic energy-pressure tensor which in general has the form 
\begin{equation}
\{T^{\mu \nu}\}=\pmatrix{{\sf P} & {\bf S}^\dagger /c 
\cr {\bf S} /c & u}.
\label{po2}
\end{equation}
In Eq.(\ref{po2}), we have (i) the Maxwell pressure tensor 
\begin{equation}
{\sf P}=\frac{1}{8\pi }\left[\big(|{\bf E}|^2+|{\bf B}|^2\big) 
{\sf 1}-2\big( {\bf E E}+{\bf B B}\big)\right],
\label{po3}
\end{equation}
(ii) the energy flux per unit time per unit area 
\begin{equation}
{\bf S}=\frac{c}{4\pi } \big[{\bf E\times B}\big],
\label{po4}
\end{equation}
and (iii) the energy density 
\begin{equation}
u=\frac{1}{8\pi } \big[|{\bf E}|^2+|{\bf B}|^2\big].
\label{po5}
\end{equation}
The radiation retardation force is thereby 
\begin{equation}
{\bf f}=\sigma \left[\left(\frac{{\bf S}-{\sf P}\cdot {\bf v}}{c}\right)
-\left(\frac{\gamma^2 {\bf v}}{c}\right)\tilde{u}\right],
\label{po6}
\end{equation}
wherein
\begin{equation}
\tilde{u}=u+
\left[\frac{{\bf v}\cdot {\sf P}\cdot {\bf v}}{c^2}\right] 
-2\left[\frac{\bf v\cdot S}{c^2}\right] .
\label{po7}
\end{equation}
This completes our theoretical derivation of classical radiation damping 
retardation forces.

\section{Classical Bremsstrahlung \label{cb}} 

Let us now consider explicit cases of electrons moving through random 
electromagnetic fields. The induced electron acceleration gives rise 
to classical electrodynamic radiation damping forces via 
bremsstrahlung. Due to the very small Thompson cross section in 
Eq.(\ref{crd4}), the strength of the classical radiation damping 
retardation forces are also very small for known cosmic electromagnetic 
fields. We seek to calculate the energy loss when the charged particle 
moves a distance \begin{math} d\ell \end{math}, i.e. 
\begin{equation}
-\frac{d{\cal E}}{d\ell} = 
-mc^2\left(\frac{d\gamma }{d\ell }\right)=|\overline{\bf f\cdot n}|,
\label{cb1}
\end{equation}
wherein \begin{math} {\bf n}={\bf v}/|{\bf v}| \end{math} is a unit vector 
in the direction of the particle velocity.
Eq.(\ref{cb1}) may computed by employing Eq.(\ref{po6}).

\subsection{Background Thermal Radiation \label{btr}}

For the cosmic background radiation, the Maxwell energy-pressure tensor 
has the form 
\begin{equation}
\{T^{\mu \nu}\}= 
\pmatrix{P_T & 0 & 0 & 0 \cr 
0 & P_T & 0 & 0 \cr
0 & 0 & P_T & 0 \cr
0 & 0 & 0 & u_T},
\label{btr1}
\end{equation}
wherein the trace condition for the electromagnetic tensor 
\begin{math} T^\mu_{\ \mu }=0  \end{math} holds true. One 
finds the relations  
\begin{eqnarray}
3P_T=u_T \ ,
\nonumber \\ 
\lambda_T=\left[\frac{\hbar c}{k_BT}\right] ,
\nonumber \\ 
u_T =\frac{\pi ^2}{15}\left[\frac{k_BT}{\lambda_T^3}\right]. 
\label{btr2}
\end{eqnarray}
The classical electrodynamic retardation force for a charged 
particle moving through black body radiation is thereby 
\begin{equation}
{\bf f}=-4\sigma P_T\gamma^2 \left[\frac{\bf v}{c}\right] ,
\label{btr3}
\end{equation}
or equivalently 
\begin{equation}
-\frac{d{\cal E}}{d\ell} =-mc^2\frac{d\gamma }{d\ell }=
4\sigma P_T\gamma^2 
\sqrt{1-\big(1/\gamma ^2 \big)}\ .
\label{btr4}
\end{equation}
In the high energy limit \begin{math} \gamma \gg 1 \end{math}, 
\begin{equation}
-\frac{d{\cal E}}{d\ell} = -mc^2 \frac{d\gamma }{d\ell}
\approx \left(\frac{4}{3}\right)\gamma ^2 \sigma u_T\ .
\label{btr5}
\end{equation}
The numerical values of the cosmic background radiation 
parameters are as follows:
\begin{eqnarray}
T\approx 2.725\ {\rm ^oK}, 
\nonumber \\ 
k_BT \approx 3.762\times 10^{-16}\ {\rm erg}
\approx 2.348\times 10^{-4}\ {\rm eV}, 
\nonumber \\ 
\lambda_T \approx 8.403 \times 10^{-2}\ {\rm cm},
\nonumber \\ 
u_T\approx 0.2604\ \left[\frac{\rm eV}{\rm cm^3}\right], 
\nonumber \\ 
\sigma_{\rm electron}u_T\approx 1.732\times 10^{-25}
\ \left[\frac{\rm eV}{\rm cm}\right], 
\nonumber \\ 
-\left[\frac{d{\cal E}}{d\ell}\right]_{\rm electron} 
\approx 2.309\times 10^{-26}
\ \left[\frac{\rm eV}{\rm cm}\right] \gamma^2, 
\nonumber \\ 
\frac{d}{d \ell }\left(\frac{1}{\gamma}\right)\approx    
4.520\times 10^{-32}\left[\frac{1}{{\rm cm}}\right],
\nonumber \\ 
\frac{d}{d \ell }\left(\frac{1}{\gamma}\right) = \frac{1}{L}\ , 
\label{btr6}
\end{eqnarray}
wherein \begin{math} \gamma \gg  1 \end{math} has been assumed. 
The length scale \begin{math} L \end{math} is enormous,   
\begin{eqnarray}
L \approx 2.214\times 10^{31}\ {\rm cm} 
\approx 7.174\times 10^9\ {\rm kpc}.
\label{btr7}
\end{eqnarray}
Thus,  
\begin{eqnarray}
\gamma (\ell ) \approx 
\frac{\gamma (0)}{1+[\gamma (0) \ell /L]}\ .
\label{btr8}
\end{eqnarray}
The classical bremsstrahlung energy losses due to the thermal 
cosmic background black body radiation can be neglected if the 
electron path length \begin{math}  \ell \end{math} obeys 
\begin{eqnarray}
\ell \ll (L/\gamma ) . 
\label{btr9}
\end{eqnarray}
The length scale associated with the size of our milky way
galaxy is 
\begin{eqnarray}
\ell_G \sim 10^{23}\ {\rm cm} \sim 30\ {\rm kpc} . 
\label{btr10}
\end{eqnarray}
For cosmic thermal radiation damping to be totally neglected on electron 
path lengths of the size of our galaxy it is sufficient that 
\begin{math} 1\ll \gamma \ll 10^7  \end{math}.

\subsection{Pomeranchuk Length \label{pl}}

In the ultra-relativistic limit \begin{math} \gamma \gg 1 \end{math}, the energy loss 
\begin{equation}
-\frac{d{\cal E}}{d\ell }=-mc^2 \frac{d \gamma }{d\ell}
\label{pl1}
\end{equation}
can be described by a general Pomeranchuk length scale 
\begin{math} L \end{math} defined by 
\begin{equation}
\frac{d}{d\ell } \left(\frac{1}{\gamma }\right)=\frac{1}{L}
\label{pl2}
\end{equation}
wherein 
\begin{equation}
\frac{1}{L}=\left[\frac{\sigma }{4\pi mc^2}\right] 
\overline{|{\bf n\times E}|^2+|{\bf n\times B}|^2
-2{\bf n}\cdot ({\bf E\times B})}
\label{pl3}
\end{equation}
and wherein \begin{math} {\bf n}  \end{math} is the unit vector in the direction of the 
velocity. The condition 
\begin{equation}
\ell \ll (L/\gamma )
\label{pl4}
\end{equation}
is sufficient for neglecting classical bremsstrahlung radiation energy losses.

\subsection{Magnetic Bremsstrahlung \label{mb}}

Another possible source of radiative damping is a domain of uniform magnetic 
field \begin{math} {\bf B} \end{math}\cite{Widrow:2002,Beck:2009}. 
When electrons move in such fields they exhibit synchrotron 
radiation due to the rotational angular frequency  
\begin{eqnarray}
{\bm \omega}_c=
-\left(\frac{e{\bf B}}{mc}\right)\frac{1}{\gamma } 
= -\left(\frac{ec{\bf B}}{\cal E}\right), 
\nonumber \\ 
\left|\frac{e}{mc}\right|\approx 
17.6086 \left[\frac{\rm Hz}{\rm 10^{-6}\ Gauss}\right].
\label{mb1}
\end{eqnarray}
The Pomeranchuk length \begin{math} L \end{math} in a uniform 
magnetic field is given by 
\begin{equation}
\frac{1}{L} = \left[\frac{\sigma }{4\pi mc^2}\right]
\left|{\bf n\times B}\right|^2 = 
\left[\frac{2\sigma u}{mc^2}\right]\sin^2 \vartheta 
\label{mb2}
\end{equation}
wherein 
\begin{equation}
u=\frac{|{\bf B}|^2}{8\pi} \ .
\label{mb3}
\end{equation}
is the energy density of the magnetic field and 
\begin{math} \vartheta  \end{math} is the angle between the velocity 
\begin{math} {\bf v} \end{math} and the magnetic field 
\begin{math} {\bf B} \end{math}. Within the milky way galaxy, the 
magnetic fieled energy density \begin{math} u\sim [{eV/cm^3}] \end{math} 
that again yields an enormous Pomeranchuk length 
\begin{equation} 
L\sim 4\times 10^{29}\left[\frac{\rm cm }{ \sin^2 \vartheta}\right]
\sim 10^7 \left[\frac{\rm kpc }{ \sin^2 \vartheta}\right] .
\label{mb4}
\end{equation} 
However, the helical paths traveled by electrons or positrons have a path 
length \begin{math} \ell  \end{math} that is more than the distance 
\begin{math} s \end{math} traveled along the magnetic field direction, i.e. 
\begin{equation}
s=\ell \cos \vartheta .
\label{mb5}
\end{equation}
The consdition that synchrotron radiation losses are neglegable then 
amounts to 
\begin{equation}
\gamma \ll \left<\left[\frac{mc^2}{2s\sigma u}\right] 
\frac{\cos \vartheta }{\sin^2 \vartheta}\right>
\label{mb6}
\end{equation}
where the average is over the size and angular orientation of the magnetic 
fields. From this point of view, electron and positrons that must travel 
distances of more than galactic proportions will have only negligable energy 
losses for \begin{math} \gamma \le 10^6 \end{math}.

\section{Conclusion \label{conc}}

Observed cosmic ray charged leptons arricing from distances of more than galactic 
proportions will exihibit bremsstrahlung energy losses and these in principle will 
have some effect on energy distributions observed employing detectors in our solar 
system. The energies \begin{math} {\cal E}=mc^2\gamma \end{math} for which 
the energy distributions will reflect those of the charged leptons emitted from 
the souce are in the range \begin{math} 1\ll \gamma \ll 10^7 \end{math}.
For distances to sources of galactic length proportions, the lepton cosmic 
ray energy must be lass than about a TeV.

\section{Acknowledgements}

J. S. would like to thank the United States National Science Foundation for support under PHY-1205845.
Y. S. would like to thank Professor Bruna Bertucci for helpful discussions.

\end{document}